\pdfoutput=1 

\documentclass{article}

\usepackage{arxiv}

\usepackage[utf8]{inputenc} 
\usepackage[T1]{fontenc}    
\usepackage{hyperref}       
\usepackage{url}            
\usepackage{booktabs}       
\usepackage{amsfonts}       
\usepackage{nicefrac}       
\usepackage{microtype}      
\usepackage{lipsum}		
\usepackage{graphicx}
\usepackage{doi}

\usepackage{balance} 
\usepackage{soul}
\usepackage{tikz}
\usepackage{multirow}
\usepackage{amsthm}

\usepackage{caption}
\usepackage{subcaption}

\usepackage{algorithm}
\usepackage{algpseudocode}
\usepackage{physics}

\title{GCS-Q: Quantum Graph Coalition Structure Generation}

\author{ \href{https://orcid.org/0000-0002-9824-7399}{\includegraphics[scale=0.06]{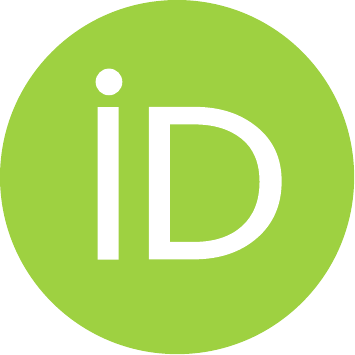}\hspace{1mm}Supreeth Mysore Venkatesh}
  \\
	Saarland University\\
	Saarbruecken, Germany\\
	\texttt{s8sumyso@stud.uni-saarland.de} \\
	\And
	\href{https://orcid.org/0000-0002-1348-250X}{\includegraphics[scale=0.06]{orcid.pdf}\hspace{1mm}Antonio Macaluso}\thanks{corresponding author} \\
	German Research Center for Artificial Intelligence (DFKI)\\
	Saarbruecken, Germany\\
	\texttt{antonio.macaluso@dfki.de} \\
	\And
        Matthias Klusch \\ 
	German Research Center for Artificial Intelligence (DFKI)\\
	Saarbruecken, Germany\\
	\texttt{matthias.klusch@dfki.de} \\
}

\renewcommand{\shorttitle}{GCS-Q: Quantum Graph Coalition Structure Generation}

\begin{document}
\maketitle

\begin{abstract}
The problem of generating an optimal coalition structure for a given coalition game of rational agents is to find a partition that maximizes their social welfare and is known to be NP-hard.
This paper proposes GCS-Q, a novel quantum-supported solution for Induced Subgraph Games (ISGs) in coalition structure generation. GCS-Q starts by considering the grand coalition as initial coalition structure and proceeds by iteratively splitting the coalitions into two nonempty subsets to obtain a coalition structure with a higher coalition value. In particular, given an $n$-agent ISG, the GCS-Q solves the optimal split problem $\mathcal{O} (n)$ times
using a quantum annealing device, 
exploring $\mathcal{O}(2^n)$ partitions at each step. 
We show that GCS-Q outperforms the currently best classical solvers with its runtime in the order of $n^2$ and an expected worst-case approximation ratio of $93\%$ on standard benchmark datasets.
\end{abstract}


\section{Problem Formulation}
\label{sec:problem formulation}

The problem of coalition structure generation (CSG) is to find a partition of $n$ rational agents into mutually disjoint coalitions such that the sum of the resulting coalition values for this coalition structure is maximized.
Given a coalition game $(A,v)$ with a set $A$ of $n$ agents and a characteristic function $v:\mathcal{P}(A) \rightarrow \mathbb{R}$ for coalition values $v(C)$ for all non-empty coalitions $C$ in $A$, the problem is to find a coalition structure $CS^*$ of $A$ that maximizes the social welfare $\sum_{C \in CS^*}v(C)$. 
An induced subgraph game (ISG) $(A,v$) is a coalition game defined by a connected, undirected, weighted graph $G(V,w)$: The set $V=\{i\}_{i \in \{1..n\}}$ of nodes in $G$ represents the set $A$
of $n$ agents, and the real-valued weights $w_{i,j}$ of edges $(i,j)\ in\ w$ denote the synergies of cooperation or joint utilities of agents $i, j \in A$ in feasible coalitions $C\subseteq A$. A coalition $C$ is feasible if and only if it induces a connected subgraph of $G$. The coalition values $v(C)$ of a feasible coalition $C$ is \begin{equation}\label{eqn:obj GSCG}
    \displaystyle v(C) =  \sum_{(i,j)\in w,~i,j\in C} w_{i,j}.
\end{equation}
For a given coalition game $(A,v)$, the coalition structures $CS$ are partitions of $A$ into mutually disjoint, feasible coalitions $C$.
The corresponding ISG is to find the optimal coalition structure $CS^*$ with maximal coalition value (or social welfare) for $(A,v)$:
\begin{equation} \label{eqn: max obj GSCG}
    CS^* = \arg \max_{CS} \sum_{C \in CS} v(C).
\end{equation}

In ISGs, the coalition values depend only on the pairwise interactions between nodes/agents. Nevertheless, the set of possible solutions is not restricted, and the problem remains NP-complete \cite{bachrach2013optimal}. 

\section{Methods}

We propose GCS-Q (\textit{Quantum-supported solution for Graph Coalition Structure generation}), a novel quantum-supported anytime approximate algorithm for solving ISG problems. 
GCS-Q starts from the grand coalition as the initial coalition structure. Then, based on the graph induced by the coalition game, it recursively performs a \textit{min-cut} to find the optimal split of the agents within each coalition. 
This approach's primary complexity source is given by finding the optimal split which is NP-hard \cite{mccormick2000min}. However, properly formulating this problem and delegating it to a quantum annealer allows for obtaining a quantum-supported algorithm capable of outperforming state-of-the-art classical and quantum solutions.

\subsection{QUBO formulation for Min-Cut Problem}\label{sec:optimal split}

Operating the \textit{min-cut} to a given graph $G(V, w)$, corresponds to finding the bipartition of $G$ with the highest possible coalition value. Here we reformulate the \textit{min-cut} problem as a QUBO (Quadratic Unconstrained Binary Optimization) in order to execute it using quantum annealing.

Consider a coalition game $(A, v)$ with underlying graph $G(V, w)$. Finding the optimal bipartition of $G$ can be formulated as a quadratic objective function of binary variables $\{x_i\}_{i=1, \dots , n}$, as follows:
\begin{equation} \label{eqn:min cut qubo}
    \begin{split}
        \arg \min_{x}\ \sum_{x_i \in C, x_j\in \overline{C}} w_{i,j} x_i (1-x_j) + {\sum w_ i x_i} = x^t W x\\
    \end{split}
\end{equation}
where $w_{i,j}$ is the weight connecting the nodes $i$ and $j$ and $w_{i}$ is a self-loop on node $i$. Associated to the solution of the objective function in Eq. \eqref{eqn:min cut qubo}, we  differentiate the vertices belonging to the subsets $C$ or $\overline{C}$ caused by the cut, as follows:
\begin{equation}
\forall x_i \in A, x_i=
    \begin{cases}
        1 & \text{if } x_i \in C\\
        0 & \text{if } x_i \in \overline{C}\\
    \end{cases}
\end{equation}
%
Given the QUBO formulation in  Eq. \eqref{eqn:min cut qubo}, we can easily define the corresponding optimization problem of an Ising Hamiltonian $\mathcal{H}$ with the assignment $x_i\rightarrow (1-Z_i)/2$, where $Z_i$ is the Pauli-$Z$ operator with eigenvalues $\pm 1$. 
Importantly, splitting the original graph into disjoint sets guarantees always obtaining a valid coalition structure for the coalition game.

\subsection{GCS-Q Algorithm}

The GCS-Q algorithm follows the strategy of hierarchical divisive clustering methods and applies it in the context of ISGs.
It starts by considering the grand coalition $g_c$ as the initial coalition structure. Then, the first step consists of splitting $g_c$ into two independent subsets $\{C, \overline{C}\}$ solving the optimal split problem on a quantum annealer. If the coalition function of the grand coalition $g_c$ (i.e., $v(g_c)$) is greater than
$v(\{C, \overline{C}\})$
(i.e., the coalition structure comprising $\{C\}$ and \{$\overline{C}\}$ separately), then the algorithm stops, and the $g_c$ is returned as the best coalition structure. Otherwise, the algorithm finds the optimal split within each coalition for the current coalition structure $\{C, \overline{C}\}$ using the quantum annealing. This process continues for each coalition until a lower coalition value is generated by performing the split.  
Notice that the additive nature of the coalition value function in Eq. \eqref{eqn:obj GSCG} for ISG problems allows acting independently on the coalitions to maximize the overall coalition value of the coalition structure.

In other terms, GCS-Q tries to split up all the coalitions of the current coalition structure into two smaller ones. 
The split is performed if and only if the generated coalitions have higher coalition values than the original. In the worst-case scenario, the hierarchy is built in $n - 1$ steps when the coalition game contains $n$ agents. A key difference between a standard divisive clustering algorithm and GCS-Q is how the partitions are determined. In the case of divisive clustering, the splitting is usually based on a distance matrix between the clusters (i.e., coalitions). This leads to a worst-case complexity (for every single step) of $n^2$.
Meanwhile, the GCS-Q considers at each step all possible partitions which amounts to $2^{n-1}$. This latter number grows exponentially fast, making this approach intractable using classical resources. However, as it will be shown in the next section, a quantum annealer can solve the optimal split problem with a runtime linear to $n$, which leads to an overall complexity in the order of $n^2$.

The pseudocode of GCS-Q is shown in Algorithm \ref{algo: CSG-Q}.

\begin{algorithm}
\caption{GCS-Q Algorithm}
\label{algo: CSG-Q}
\raggedright \textbf{Input: }{Coalition game $(A,v)$, with underlying graph $G(A, w)$ where $|A|= n$ and ${w}:A \cross A \to \mathbb{R}$} \\
\raggedright \textbf{Output: }{Optimal coalition structure $CS^*$}
\begin{algorithmic}[1] \label{alg: CSG-Q}
\State $CS^* \leftarrow \{\}$ \Comment{initialize $CS^*$ with empty list}
\State $queue \leftarrow g_c$ \Comment{initialize $queue$ with grand coalition $g_c$}
\While {$queue \neq \emptyset$}
    \State $S \leftarrow queue.pop $ \Comment{Fetch a coalition from $queue$}
    \State \Comment{Optimal Split problem} 
    \State  Create a weight matrix $W$ from edges in $S$  \Comment{Eq. \eqref{eqn:min cut qubo}}
    \State  Define the Ising Hamiltonian $\mathcal{H}$ for $W$ 
    \State  Solve Ising Hamiltonian $\mathcal{H}$ on a quantum annealer
    \State  Decode binary string to get $C$ and $\overline{C}$
    \State \Comment{End Optimal Split problem}
    \If{$\overline{C} = \emptyset $} \Comment{no splits of $S$ is better than $S$}
        \State \textbf{add} $C$ to $CS^*$
    \Else \Comment{try further splitting $C$ and $\overline{C}$}
        \State \hskip0.5em \textbf{add} $C$ to $queue$
        \State \hskip0.5em \textbf{add} $\overline{C}$ to $queue$
    \EndIf
\EndWhile
\State \textbf{Return} $CS^*$
\end{algorithmic}
\end{algorithm}

To the best of our knowledge, only one approximate solver can find a solution for an ISG problem with a fully connected graph in polynomial time, the C-Link algorithm \cite{farinelli2013c}. This approach is based on agglomerative clustering and makes decisions by considering the local patterns without ever considering the global patterns in the coalition game. 
On the contrary, the top-down approach proposed by GCS-Q examines an exponentially large number of partitions at each step. Thus, GCS-Q explores a larger portion of the solution space, potentially leading to a better quality of solutions.

\section{Evaluation}

\begin{figure}[ht]
\centering
         \includegraphics[scale=.25]{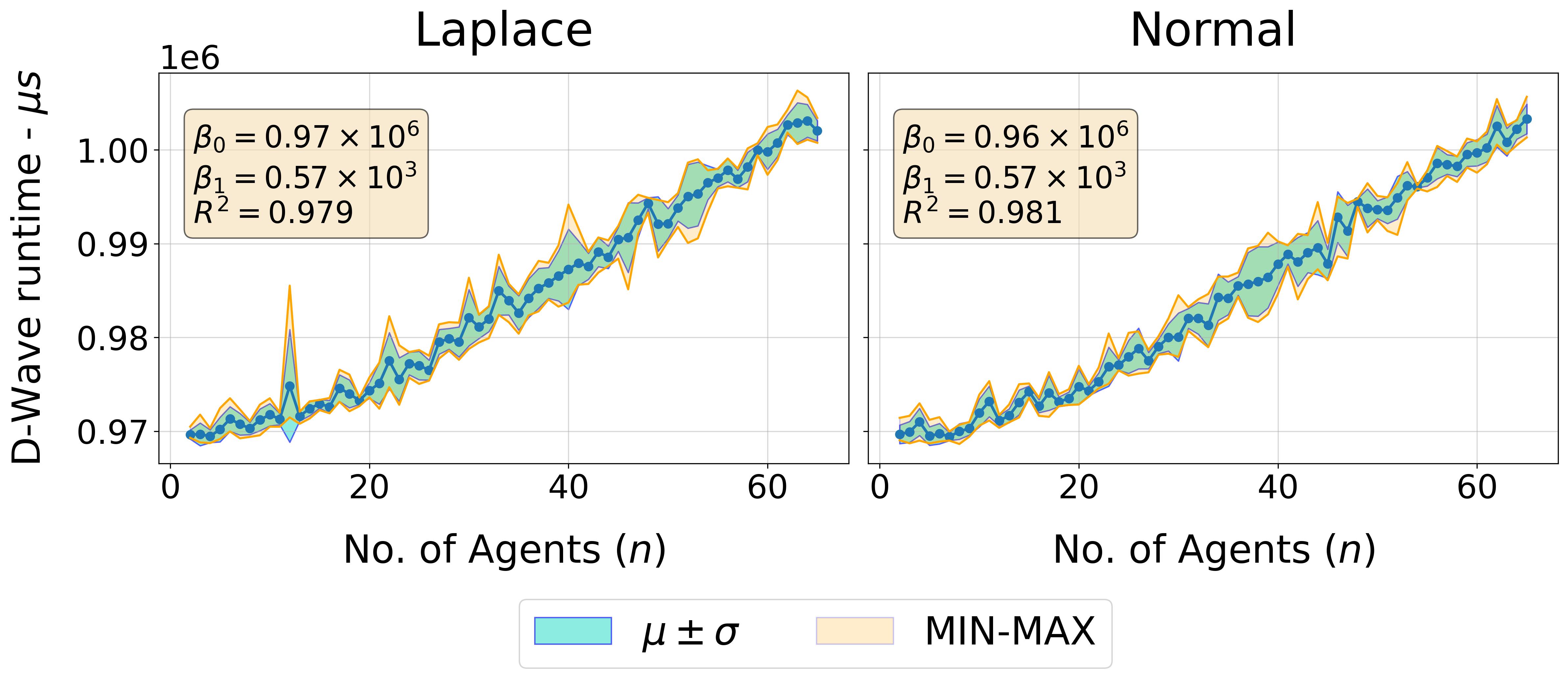}
     \caption{
     The runtime of the D-Wave 2000Q when solving the optimal split problem. For each number of agents, the same QUBO problem is solved $5$ times. The blue line is the average runtime of the $5$ experiments. The yellow shaded area represents the maximum and minimum runtimes. The green shaded area is calculated considering the mean and the standard deviation of the runtimes for each problem instance. The runtimes are reported in microseconds ($\mu s)$.}
    \label{fig: quantum optimal split}
\end{figure}

In this section, we evaluate the GCS-Q in terms of runtime and  approximation ratio on synthetic datasets, running experiments on a real quantum annealer, the D-Wave 2000Q. 
In order to estimate the runtime of the quantum annealing in solving the optimal split problem (sec. \ref{sec:optimal split}), two sets of ISGs are generated by sampling the interaction score associated with the pairs of agents using \textit{Laplace} and \textit{Normal} distributions\footnote{ $\mathcal{L}(\mu=0, b=5)$, \hspace{.3em} $\mathcal{N}(\mu=0, \sigma=5)$}.  The runtime for both distributions is reported in Figure \ref{fig: quantum optimal split}.

The order of growth of the runtime when increasing the number of agents is linear. To confirm this hypothesis, we estimate a linear regression and calculate the coefficient of determination $R^2$, which is equal to $1$ in the case of a perfect deterministic linear function between the number of agents $n$ and the runtime. A value of $97\%$ for both distributions indicates that the relationship increases almost perfectly linearly in the order of $\beta_1$. Thus, we can conclude that the average-case complexity of the quantum annealing solution for solving the optimal split problem  $\Theta(n)$.

\begin{figure}
\center
\includegraphics[scale=.22]{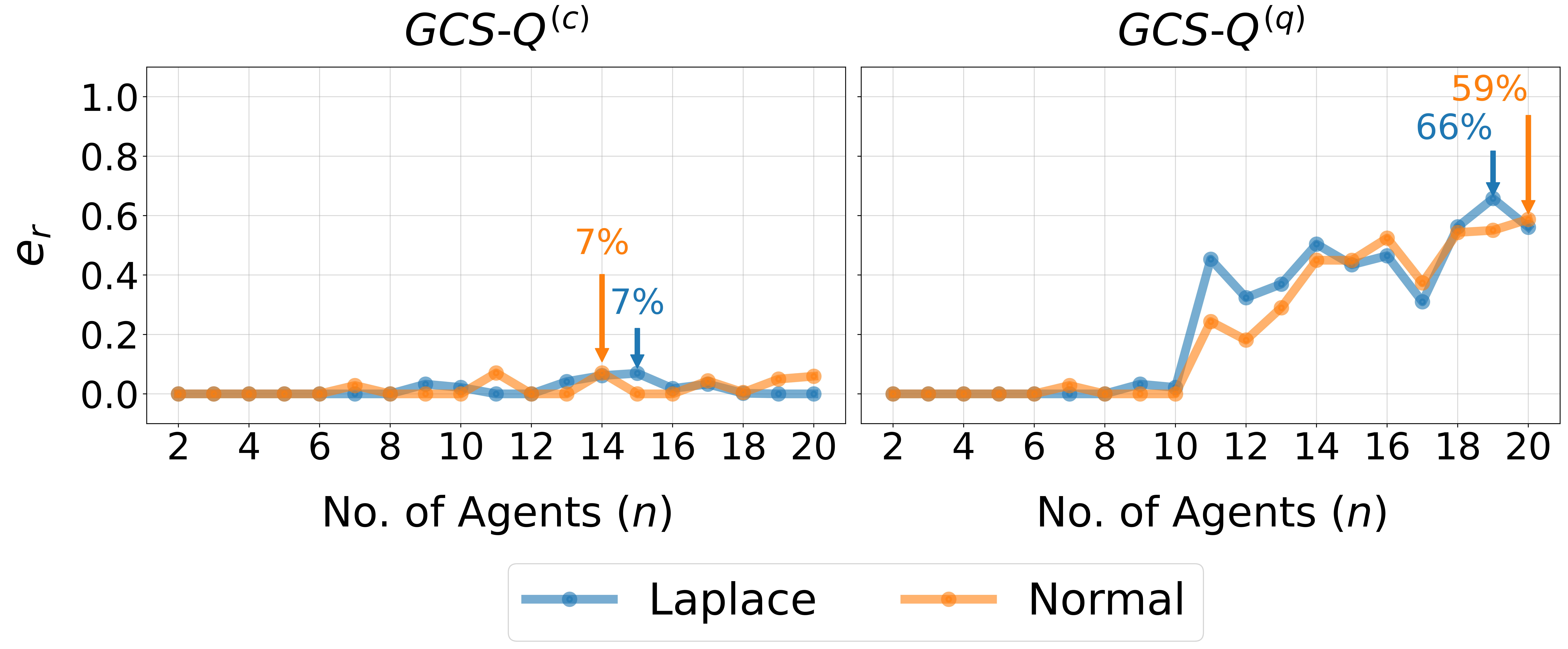}
\caption{Assessment of GSC-Q quality in terms of approximation error  calculated on the two benchmark datasets. The experiments are limited up to $20$ agents since IDP has space complexity of $\mathcal{O}(2^n)$ and time complexity of $\mathcal{O}(3^n)$, which makes most machines to go out of resources. 
}
\label{fig:dcq classical error}
\end{figure}

Two sets of coalition games are generated based on the two distributions mentioned above to assess the quality of GCS-Q. Thus, we implement the Algorithm \ref{algo: CSG-Q} into two variants: GCS-Q$^{(c)}$ solves the QUBO at each step using a classical QUBO solver, which always returns the exact solution. While GCS-Q$^{(q)}$ uses a D-Wave 2000Q.  We compare the solutions of these two approaches for coalition games up to $20$ agents with the optimal solution obtained by running the IDP \cite{rahwan2008improved} algorithm on the same problem instances.
Results are shown in Figure \ref{fig:dcq classical error}.

For both distributions, the GCS-Q$^{(c)}$ has the largest error observed equally to $7\%$. While the GCS-Q$^{(q)}$ produces the expected solution of the Algorithm \ref{algo: CSG-Q} only for coalition games of size up to $10$. The approximation error suddenly increases for problems with more than $10$ agents.
The decrease in performance is due to the poor quality of the solution provided by the quantum annealer. In fact, with $n \geq 11$, the error is cascaded through further executions of the algorithm, and the final solution attained for the ISG problem is far from optimality.

\subsection{Performance analysis} \label{sec:dcq performance analysis}

In this section, we compare the runtime of GCS-Q with state-of-the-art classical and quantum solutions when considering an ISG problem with an underlying fully connected graph. Since ISGs can be viewed as a particular case of standard CSG, we include general solvers in the comparative analysis.

The state-of-the-art classical exact solvers for CSG problems are represented by methods based on IDP \cite{rahwan2008improved}, such as BOSS\cite{changder2021boss} and DyCE\cite{voice2012dyce}, with a worst-case time complexity of $\mathcal{O}(3^n)$.
For ISGs, the CFSS \cite{bistaffa2014anytime} and KGC \cite{bistaffa2021efficient} algorithms  have shown good results with sparse graphs, but the worst-case complexity for a complete graph still remains $\mathcal{O}(n^n)$. 
The best approximate solution in terms of the runtime is C-link \cite{farinelli2013c}, with scales as $\mathcal{O}(n^3)$.  

\begin{figure}[H]
\small
\center
\includegraphics[scale=.45]{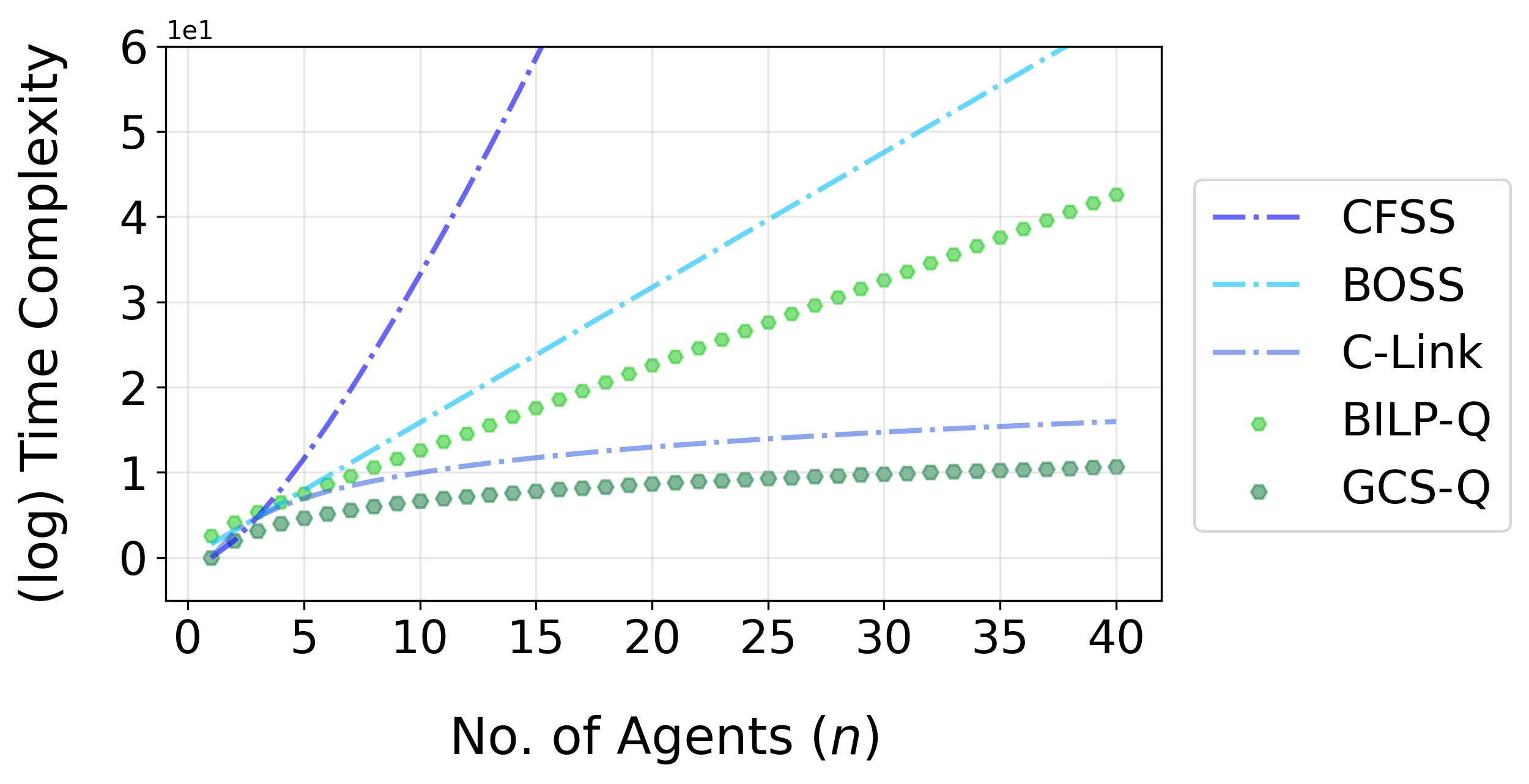}
\vspace{-1em}
\caption{Cost complexity as a function of the number of agents $n$. Classical solutions are indicated with blue lines, while quantum solutions in green. }
\label{fig:performance analysis}
\end{figure}

BILP-Q\cite{venkatesh2022bilp} is the first quantum-supported solution with an average runtime in the order of $2^n$ on quantum annealing.
A graphical comparison of classical and quantum solutions is provided in Figure \ref{fig:performance analysis}.

The ability of GCS-Q to be executed with a runtime quadratic in the number of agents makes it the best solver for ISG problems.

\section{Discussion}

The proposed GCS-Q represents the best solver in terms of runtime for ISG problems. However, the performance of the D-Wave 2000Q causes obtaining poor quality solutions for problems with more than ten agents.
The limits of the D-Wave 2000Q have already been emphasized for specific optimization tasks\cite{willsch2022benchmarking}. Nonetheless, the latest generation of D-Wave, named \textit{Advantage}, outperformed D-Wave 2000Q for any problem size while solving larger problem instances. 
In some cases, not only the \textit{Advantage} system can find better-quality solutions than the D-Wave 2000Q, but also obtains same quality solutions faster \cite{tasseff2022emerging}\footnote{A comparison between \textit{Advantage} with the D-Wave 2000Q is reported \cite{mcgeoch2020d}
}.
Nevertheless,  several optimization strategies are still possible on the D-Wave 2000Q. For instance, the adoption of different embedding strategies or using hybrid quantum annealing that can deliver better quality solutions at the cost of worsening the runtime (for more details, see D-Wave documentation\footnote{\href{https://docs.dwavesys.com/docs/latest/index.html}{https://docs.dwavesys.com/docs/latest/index.html}}).

Compared with other approximate solvers, such as C-Link \cite{farinelli2013c}, GCS-Q examines a larger portion of the solution space, potentially leading to a better approximation ratio. In fact, C-link relies on hierarchical agglomerative clustering where only local interactions are adopted for merging coalitions. In other terms, given a set of $n$ coalitions, C-link explores the distance matrix between coalitions and merges only the pair of coalitions with the minimum distance ($n^2$ possibilities). On the contrary, GCS-Q examines all possible bipartition of a given coalition exploring $\mathcal{O}(2^n)$ possibilities at each step. This implies the total number of solutions analyzed to be (worst-case) in the order of $\mathcal{O}(n2^n)$ and thus significantly higher than the number of solutions investigated by C-Link, which is $\mathcal{O}(n^3)$.

With respect to other quantum solutions, such as BILP-Q \cite{venkatesh2022bilp}, GCS-Q delivers a better qubit complexity. Although being an exact solver, BILP-Q requires $2^n$ logical qubits to be executed, which is a substantial limitation considering near-term quantum computers. In comparison, GCS-Q can solve an $n$-agent ISG using $n$ logical qubits.

\section{Conclusions}

We proposed GCS-Q, a novel quantum-supported solution for induced subgraph games in coalition structure generation.
The key idea is to partition the graph underlying the coalition game into two subsets to obtain a coalition structure with a better coalition value at each step. By delegating the task of finding the optimal split to a quantum annealer, we obtain a solver capable of running faster than the state-of-the-art solutions (quantum and classical). 

In addition, we provided a practical implementation of GCS-Q and evaluated its performance on standard benchmark datasets. Specifically, we generated two sets of ISGs with fully connected graphs by sampling the edge weights from two distributions (\textit{Laplace} and \textit{Normal}).

When calculating the quality of the solutions, the performance of GCS-Q deteriorates due to the limitations of the quantum hardware in use. 
For this reason, the main challenge to tackle in the near future is the investigation of alternative embedding strategies or the adoption of hybrid quantum-classical solvers proposed by D-Wave.

Another natural follow-up is the execution of GCS-Q on better quantum hardware, such as the D-Wave \textit{Advantage}. This latest generation of quantum devices has outperformed the D-Wave 2000Q in terms of both quality of solutions and runtime in several combinatorial optimization problems.

In conclusion, we showed the feasibility and the benefit of adopting quantum computation in multi-agent systems. Thanks to this work, we believe that GCS-Q can represent the first quantum-supported solutions suitable for solving real-world AI problems on near-term quantum technology.

\section*{Acknowledgments}

This work has been funded by the German Ministry for Education and Research (BMB+F) in the project QAI2-QAICO under grant 13N15586.





\bibliographystyle{unsrt}
\bibliography{references}  






\end{document}